\documentclass{emulateapj}
\RequirePackage{lineno}
\usepackage{amssymb}
\usepackage{amsmath}
\usepackage{mathrsfs}
\usepackage{natbib}
\usepackage[colorlinks, linkcolor=blue, urlcolor=blue, citecolor=blue]{hyperref}
\usepackage{array}
\usepackage{float}
\usepackage{graphicx}
\usepackage{subfigure}
\usepackage{color}
\usepackage[all]{hypcap}
\usepackage[toc,page]{appendix}

\def\beq{\begin{equation}}
\def\enq{\end{equation}}
\def\bea{\begin{eqnarray}}
\def\ena{\end{eqnarray}}

\begin{document}

\title{Constraining Parameters in Pulsar Models of Repeating FRB 121102 with High-Energy Follow-up Observations}
\author{Di Xiao\altaffilmark{1,2} and Zi-Gao Dai\altaffilmark{1,2}}
\affil{\altaffilmark{1}School of Astronomy and Space Science, Nanjing University, Nanjing 210093, China; dzg@nju.edu.cn}
\affil{\altaffilmark{2}Key Laboratory of Modern Astronomy and Astrophysics (Nanjing University), Ministry of Education, China}

\begin{abstract}
Recently, a precise (sub-arcsecond) localization of the repeating fast radio burst (FRB) 121102 has led to the discovery of persistent radio and optical counterparts, the identification of a host dwarf galaxy at a redshift of $z=0.193$, and several campaigns of searches for higher-frequency counterparts, which gave only upper limits on the emission flux. Although the origin of FRBs remains unknown, most of the existing theoretical models are associated with pulsars, or more specifically, magnetars. In this paper, we explore persistent high-energy emission from a rapidly rotating highly magnetized pulsar associated with FRB 121102 if internal gradual magnetic dissipation occurs in the pulsar wind. We find that the efficiency of converting the spin-down luminosity to the high-energy (e.g., X-ray) luminosity is generally much smaller than unity, even for a millisecond magnetar. This provides an explanation for the non-detection of high-energy counterparts to FRB 121102. We further constrain the spin period and surface magnetic field strength of the pulsar with the current high-energy observations. In addition, we compare our results with the constraints given by the other methods in previous works and would expect to apply our new method to some other open issues in the future.
\end{abstract}


\section{Introduction}
The origin of fast radio bursts (FRBs) has been under intense debate since their discovery ten years ago \citep{lor07, kea12, tho13, bur14, spi14, Spitler2016, cha15, mas15, pet15, rav16, Caleb2017}. Most of the 23 FRBs detected so far appear to be non-repeating, and various origin models for these kind of events have been proposed, such as collapse of supra-massive neutron stars to black holes \citep{fal14,zha14}, mergers of binary white dwarfs \citep{kas13} or binary neutron stars \citep{tot13, wan16}, or charged black holes \citep{Liu2016, zha16}, and so on \citep[for a review of observations and physical models see][]{Katz2016a}.

However, the discovery of the only repeating FRB 121102 shed new light on its origin, since the catastrophic event scenarios are not suitable for it \citep{Spitler2016}. The non-catastrophic models include giant flares from a magnetar \citep{pop13, kul14, kat16b},\footnote{Although this model is challenged by the non-detection of an expected bright radio burst during the 2004 December 27 giant gamma-ray flare of the Galactic magnetar SGR 1806-20, it is still possible to reconcile the theory with observations \citep{ten16}.} giant pulses from a young pulsar \citep{Connor2016,Cordes2016,lyu16}, pulsar lightning \citep{katz17}, repeating collisions of a neutron star, an asteroid belt around another star \citep{dai16,Bagchi2017}, and accretion in a neutron star-white dwarf binary \citep{Gu2016}. Among these models, a rapidly rotating highly magnetized neutron star is the astrophysical object referred to most frequently. Moreover, the properties of the host dwarf galaxy of FRB 121102, which are consistent with those of long-duration gamma-ray bursts (GRBs) and hydrogen-poor superluminous supernovae (SLSNe), suggest the possibility that the repeating bursts originate from a young millisecond magnetar \citep{Metzger2017}. This possibility is further  supported by the location of FRB 121102 within a bright star-forming region \citep{Bassa2017}. In addition, based on the recently discovered persistent radio source associated with FRB 121102 and the redshift of the host galaxy \citep{Chatterjee2017,Marcote2017,Tendulkar2017}, some constraints on the pulsar were widely discussed \citep[e.g.][]{Bel2017, Cao2017, Dai2017, Kashiyama2017,Lyutikov2017}, but they were relaxed under the assumption that FRBs are wandering narrow beams \citep{katz17b}. To our knowledge, in addition to FRBs, a pulsar can exhibit some observational signals in other wavelengths. Thus, searching for high-energy counterparts to FRB 121102 will possibly give us some hints about the central pulsar.

A pulsar is likely to generate an ultra-relativistic wind, and there are some observational signatures for such a wind. The measured radio spectrum of the Crab Nebula is naturally explained if a wind with a Lorentz factor of $\sim10^4$ from the Crab pulsar is introduced \citep[e.g.][]{ato99}. The wind from a rapidly rotating highly magnetized pulsar is expected to be Poynting-flux-dominated \citep{cor90, spr01} or alternatively turns into electron-positron pairs dominated above a certain radius, and then, even powering a GRB afterglow is possible \citep[e.g.][]{dai98a,dai98b,dai04, cio15, rez15}. The gradual dissipation of magnetic energy via reconnection is able to accelerate electrons and then produce radiation \citep{spr01, dre02a, dre02b, gia05, gia06, gia08, met11, gia12, sir14, ben14, kag15, sir15}.\footnote{This kind of gradual dissipation of magnetic energy via reconnection in these references is different from an abrupt and violent dissipation process arising from colliding shells in the internal-collision-induced magnetic reconnection and turbulence model proposed by \citet{Zhang11}. This model can account for the main properties of GRBs themselves.} Recently, \citet{ben17} found that this emission could be significant in the X-ray/gamma-ray band, which motivates us to constrain the parameters of the pulsar, especially its spin period and surface magnetic field strength, with the non-detection of high-energy counterparts to FRB 121102.

The observational results that we refer to mainly include three upper limits given by different instruments for different working bands, and are summarized below. A deep search for X-ray sources by {\em XMM-Newton/Chandra} placed a $5\sigma$ upper limit of $4.0\times10^{-15}\,\rm erg\,cm^{-2}\,s^{-1}$ on the $0.5-6\,\rm keV$ flux \citep{sch17}. The $5\sigma$ flux upper limit by {\em Fermi-GBM} is $1.0\times10^{-7}\,\rm erg\,cm^{-2}\,s^{-1}$ \citep{sch16, you16}. In addition, an energy flux upper limit of $4.0\times10^{-12}\,\rm erg\,cm^{-2}\,s^{-1}$ was obtained over the eight-year span of {\em Fermi-LAT} \citep{zhabb17, xi17}.

The paper is organized as follows. We introduce an internal gradual magnetic dissipation model (abbreviated as the IGMD model hereafter) of a wind from a rapidly rotating highly magnetized pulsar and predict an emission from the wind in Section 2. Then we calculate the radiation efficiency and constrain the spin period and surface magnetic field strength in Section 3. In Section 4 we provide a summary and compare with previous works, and also discuss an implication for future works.

\section{Emission from a Pulsar Wind}
 An ultra-relativistic wind from a rapidly rotating highly magnetized pulsar is initially Poynting-flux-dominated \citep{cor90}, and its magnetic energy can be converted to thermal emission and bulk kinetic energy of the wind via internal gradual magnetic dissipation due to reconnection in the IGMD model \citep{spr01, dre02a, dre02b, gia05}. In addition, we also expect to observe non-thermal synchrotron emission from the electrons accelerated by magnetic reconnection \citep{ben14, sir14, kag15}. At a given radius, the Poynting-flux luminosity could be written as \citep{gia05, ben17}
\beq
L_B=c\frac{(rB)^2}{4\pi}=L_{\rm sd}\left[1-\frac{\Gamma(r)}{\Gamma_{\rm sat}}\right],
\enq
where $B$ and $\Gamma(r)$ are the magnetic field strength and Lorentz factor of the wind at radius $r$ respectively. The energy injection luminosity of the wind is assumed to be the spin-down luminosity $L_{\rm sd}$ and $\Gamma_{\rm sat}$ is the bulk Lorentz factor of the wind at the saturation radius given by $r_{\rm sat}=\lambda\Gamma_{\rm sat}^2/(6\epsilon)=1.7\times10^{15}\Gamma_{\rm sat,4}^2(\lambda/\epsilon)_8\,\rm cm$ \citep{ben17}, where $\lambda\sim cP=3\times10^7P_{-3}\,\rm cm$ is the wavelength of the magnetic field in the striped wind configuration \citep{cor90, spr01, dre02a, dre02b} and $\epsilon\sim0.1-0.25$ is the ratio of reconnection velocity to the speed of light \citep{lyu05, guo15, liu15}. Throughout this work, we use the notation $Q=10^xQ_x$ in cgs units. Since the comoving temperature decreases as $T^{\prime}\propto r^{-7/9}$, the thermal luminosity decreases as $L_{\rm th}(r)\propto r^{-4/9}$ \citep{gia05}, substituting the energy dissipation rate $d\dot{E}=-(dL_B/dr)dr$; then the total thermal photospheric luminosity can be obtained by integrating from the initially launching radius to the photospheric radius $r_{\rm ph}$ \citep{gia05, ben17},
\bea
L_{\rm ph}&=&\int_0^{r_{\rm ph}}\frac{1}{2}\left(\frac{r}{r_{\rm ph}}\right)^{4/9}d\dot{E}\nonumber\\
&=&2.6\times10^{47}L_{\rm sd,50}^{6/5}\Gamma_{\rm sat,4}^{-1}\left(\frac{\lambda}{\epsilon}\right)_8^{-1/5}\,\rm erg\,s^{-1}\,sr^{-1},
\ena
with the temperature being\footnote{Note that the coefficient and indexes derived in Equation (\ref{Tph})
are different from those of \cite{ben17}.}
\beq
T_{\rm ph} = 95L_{\rm sd,50}^{1/10}\Gamma_{\rm sat,4}^{1/4}\left(\lambda\over \epsilon\right)_8^{-7/20}\,{\rm keV}\label{Tph},
\enq
where $r_{\rm ph}$ can be obtained by setting the Thomson scattering depth $\tau(r_{\rm ph})=1$, which gives
$r_{\rm ph}=3.0\times10^9L_{\rm sd,50}^{3/5}\Gamma_{\rm sat,4}^{-1}(\lambda/\epsilon)_8^{2/5}\,\rm cm$ \citep{ben17}.

Furthermore, in order to obtain the synchrotron spectrum, we need to calculate the relevant break frequencies. The acceleration timescale due to magnetic reconnection is $t_{\rm acc}=(\gamma_em_ec^2)/(q\epsilon B^{\prime}c)$ \citep{gia10}, where $q$ is the electron charge and $B^{\prime}$ is the comoving magnetic field strength of the wind, while the synchrotron cooling timescale is $t_{\rm syn}=(6\pi m_ec)/(\sigma_TB^{\prime2}\gamma_e)$, where $\sigma_T$ is the Thomson scattering cross-section. Thus, letting $t_{\rm acc}=t_{\rm syn}$ gives the maximum Lorentz factor of electrons,
\beq
\gamma_{\max}=\left(\frac{6\pi q\epsilon}{\sigma_TB^{\prime}}\right)^{1/2}.
\enq
Correspondingly, the maximum synchrotron frequency in the observer's rest-frame is
\beq
\nu_{\max}=\frac{1}{1+z}\Gamma\gamma_{\max}^2\frac{qB^{\prime}}{2\pi m_ec}.
\enq
The minimum Lorentz factor $\gamma_m$ depends on the spectrum of electrons. PIC simulations suggest that the accelerated electrons, through reconnection, could have a power-law distribution with an index $p$ \citep{sir14, guo15, kag15, wer16}, where $p=4\sigma^{-0.3}$ is adopted in accordance with previous numerical results, where $\sigma$ is the magnetization parameter. If $p<2$, we can simply assume $\gamma_m\simeq 1$. For $p>2$, the minimum Lorentz factor is \citep{ben17}
\beq
\gamma_m=\frac{p-2}{p-1}\frac{\epsilon_e}{2\xi}\sigma(r)\frac{m_p}{m_e},
\enq
where $\epsilon_e\sim 0.2$ is the fraction of the dissipated energy per electron and $\xi\simeq 0.2$ is the fraction of the electrons accelerated in the reconnection sites \citep{sir15}. The typical synchrotron frequency is then
\beq
\nu_m=\frac{1}{1+z}\Gamma\gamma_m^2\frac{qB^{\prime}}{2\pi m_ec}.
\enq
In addition, the cooling frequency is \citep{sar98}
\beq
\nu_c=\frac{1}{1+z}\frac{72\pi em_ec^3\Gamma^3}{\sigma_T{B^{\prime}}^3r^2}.
\enq
Letting $\nu_m=\nu_c$, we can obtain the radius $r_{\rm tr}$ at which the transition from fast cooling to slow cooling happens.

For $r_{\rm ph}<r\leq r_{\rm tr}$ and no synchrotron self-absorption (SSA), the electrons are in the fast-cooling regime for which the spectrum is \citep{sar98}
\beq
L_\nu^{\rm syn}=\begin{cases}
L_{\nu,\max}^{\rm syn}(\nu/\nu_c)^{1/3} & \text{if}\,\,\nu<\nu_c, \\
L_{\nu,\max}^{\rm syn}(\nu/\nu_c)^{-1/2} & \text{if}\,\,\nu_c<\nu<\nu_m, \\
L_{\nu,\max}^{\rm syn}(\nu_m/\nu_c)^{-1/2}(\nu/\nu_m)^{-p/2} & \text{if}\,\,\nu_m<\nu<\nu_{\max},
\end{cases}
\enq
where
\begin{equation}
L_{\nu,\max}^{\rm syn}=(1+z)\frac{m_ec^2\sigma_T\Gamma B^{\prime}N_e(r)}{3q},
\end{equation}
with $N_e(r)$ being the total number of emitting electrons in the wind at $r$.
For $r_{\rm tr}\leq r\leq r_{\rm sat}$, the electrons turn into the slow-cooling regime and the spectrum becomes \citep{sar98}
\beq
L_\nu^{\rm syn}=\begin{cases}
L_{\nu,\max}^{\rm syn}(\nu/\nu_m)^{1/3} & \text{if}\,\,\nu<\nu_m, \\
L_{\nu,\max}^{\rm syn}(\nu/\nu_m)^{-(p-1)/2} & \text{if}\,\,\nu_m<\nu<\nu_c, \\
L_{\nu,\max}^{\rm syn}(\nu_c/\nu_m)^{-(p-1)/2}(\nu/\nu_c)^{-p/2} & \text{if}\,\,\nu_c<\nu<\nu_{\max}.
\end{cases}
\enq

However, the SSA effect might play a role, and its frequency $\nu_a$ and corresponding electron Lorentz factor $\gamma_a$ satisfy
\beq
\frac{2\nu_a^2}{c^2}\gamma_a\Gamma m_ec^2\frac{\pi r^2}{\Gamma^2}=\frac{L_{\nu_a}^{\rm syn}}{(1+z)^3},
\enq
where
\beq
\nu_a=\frac{1}{1+z}\Gamma\gamma_a^2\frac{qB^{\prime}}{2\pi m_ec}.
\enq
At $r_{\rm ph}$, usually $\nu_a>\nu_c$, and then at a radius $r_{\rm cr}$, $\nu_a$ crosses $\nu_c$.
Since the spectrum below $\nu_a$ is $L_\nu\propto\nu^{11/8}$ \citep{gra02}, the whole synchrotron spectrum can be written as follows. Initially, for $r_{\rm ph}<r\leq r_{\rm cr}$,
\beq
L_\nu^{\rm syn}=\begin{cases}
L_{\nu_a}^{\rm syn}(\nu/\nu_a)^{11/8} & \text{if}\,\,\nu<\nu_a, \\
L_{\nu_a}^{\rm syn}(\nu/\nu_a)^{-1/2} & \text{if}\,\,\nu_a<\nu<\nu_m, \\
L_{\nu_a}^{\rm syn}(\nu_m/\nu_a)^{-1/2}(\nu/\nu_m)^{-p/2} & \text{if}\,\,\nu_m<\nu<\nu_{\max}.
\end{cases}
\enq
Furthermore, for $r_{\rm cr}\leq r\leq r_{\rm tr}$,
\beq
L_\nu^{\rm syn}=\begin{cases}
L_{\nu_a}^{\rm syn}(\nu/\nu_a)^{11/8} & \text{if}\,\,\nu<\nu_a, \\
L_{\nu_a}^{\rm syn}(\nu/\nu_a)^{1/3}  & \text{if}\,\,\nu_a<\nu<\nu_c, \\
L_{\nu_a}^{\rm syn}(\nu_c/\nu_a)^{1/3}\\ \,\,\times (\nu/\nu_c)^{-1/2} & \text{if}\,\,\nu_c<\nu<\nu_m, \\
L_{\nu_a}^{\rm syn}(\nu_c/\nu_a)^{1/3}\\ \,\,\times (\nu_m/\nu_c)^{-1/2}(\nu/\nu_m)^{-p/2} & \text{if}\,\,\nu_m<\nu<\nu_{\max}.
\end{cases}
\enq
Lastly, for $r_{\rm tr}\leq r\leq r_{\rm sat}$,
\beq
L_\nu^{\rm syn}=\begin{cases}
L_{\nu_a}^{\rm syn}(\nu/\nu_a)^{11/8} & \text{if}\,\,\nu<\nu_a, \\
L_{\nu_a}^{\rm syn}(\nu/\nu_a)^{1/3}  & \text{if}\,\,\nu_a<\nu<\nu_m, \\
L_{\nu_a}^{\rm syn}(\nu_m/\nu_a)^{1/3}\\ \,\,\times (\nu/\nu_m)^{-(p-1)/2} & \text{if}\,\,\nu_m<\nu<\nu_c, \\
L_{\nu_a}^{\rm syn}(\nu_m/\nu_a)^{1/3}\\ \,\,\times (\nu_c/\nu_m)^{-(p-1)/2}(\nu/\nu_c)^{-p/2} & \text{if}\,\,\nu_c<\nu<\nu_{\max}.
\end{cases}
\enq

The non-thermal synchrotron spectrum can be obtained by integrating the above expressions from the photospheric radius to the saturation radius. Now we can plot in Figure 1 the radiation spectrum of the pulsar wind, assuming the spin period $P=5\,\rm ms$ and the surface magnetic field strength $B_s=10^{14}\,\rm G$. The distance in our calculations has been implicitly assumed to be at redshift $z=0.193$, which corresponds to a luminosity distance of $972\,\rm Mpc$. For different parameter sets, different cases are named in the form of ``P$x$Bs$y$'', with $x$ denoting the spin period in ms and $y$ denoting the logarithm of the magnetic field strength in Gauss, where we are discussing the P5Bs14 case. The spin-down luminosity is then
\beq
L_{\rm sd}=L_0\left(1+\frac{t}{T_{\rm sd}}\right)^{-2},
\enq
where the present spin-down timescale $T_{\rm sd}=2\times10^3I_{45}B_{s,15}^{-2}P_{-3}^2R_6^{-6}\,\rm s$ with $I$ being the moment of inertia and $R$ being the stellar radius, so that $t<T_{\rm sd}$ and $L_{\rm sd}\simeq L_0=3.8\times10^{49}B_{s,15}^2P_{-3}^{-4}R_6^6\,\rm erg\,s^{-1}$ is a good approximation. The total spectrum (solid line) in Figure 1 consists of thermal (dotted line) and non-thermal (dashed line) components, and black upper limits are given by high-energy observations. We can see that the X-ray observations give the tightest constraint. Note that P5Bs14 is a nominal parameter set. If more optimistic parameters (e.g. P1Bs15) are taken, the X-ray flux of the wind is higher than the upper limit shown in Figure 1, so the X-ray emission should be detected by {\em XMM-Newton/Chandra}. In other words, the IGMD model will potentially be tested if a high-energy counterpart to any FRB is detected in the future.

\begin{figure}
\begin{center}
\includegraphics[scale=0.4]{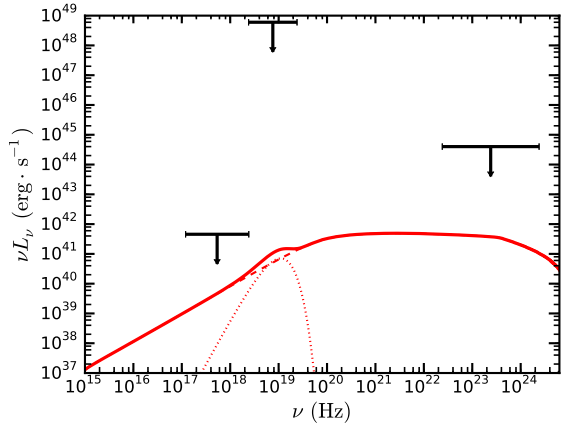}
\caption{Radiation spectrum of a pulsar wind with internal gradual magnetic energy dissipation, assuming a spin period $P=5\,\rm ms$ and surface magnetic field strength $B_s=10^{14}\,\rm G$. The total spectrum is represented by the solid line, which consists of the thermal component (dotted line) and the synchrotron emission (dashed line). Three upper limits in black are given by {\em XMM-Newton/Chandra, Fermi-GBM} and {\em Fermi-LAT} respectively.}
\end{center}
\end{figure}

The efficiency of converting the spin-down luminosity to the X-ray emission observed by {\em XMM-Newton/Chandra} can be calculated by
\beq
\eta_X\equiv\frac{\int_{\rm 0.5\,keV}^{\rm 6\,keV}(L_\nu^{\rm ph}+L_\nu^{\rm syn})d\nu}{L_{\rm sd}}\label{etax}.
\enq
For the P5Bs14 case here, we find $\eta_X=4.7\times10^{-4}$.

\section{Constraining Pulsar Parameters}
\label{sec:cons}
In order to be consistent with the upper limits given by {\em XMM-Newton/Chandra}, a solid requirement could be written as
\beq
\eta_XL_{\rm sd}\lesssim L_{X,\rm lim},
\label{eq:limit}
\enq
where $L_{X,\rm lim}$ is given by the observations (assuming $z=0.193$). Therefore, we need to find a relationship between the X-ray efficiency and the spin-down luminosity. We choose three different spin periods ($P=1,\,3$ and $5\,\rm ms$) and three different field strengths ($B_s=10^{14},\,10^{15}$, and $10^{16}\,\rm G$), so that there are nine cases. The nine efficiencies are obtained in the same way as described in the previous section. In Figure 2 we plot the dependence of $\eta_X$ on $L_{\rm sd}$ and fit it with a polynomial. The best fit is expressed as
\beq
\log\eta_X = -0.007794(\log L_{\rm sd})^2 + 0.9829\log L_{\rm sd}-31.71\label{eta}.
\enq
Substituting into the requirement (\ref{eq:limit}), we can get the critical spin-down luminosity $L_{\rm sd,cr}=8.68\times10^{44}\,\rm erg\,s^{-1}$, which is obtained by letting $\eta_XL_{\rm sd}=L_{X,\rm lim}$. Therefore, the requirement $L_{\rm sd}<L_{\rm sd,cr}$ gives a constraint on the spin period and field strength of the pulsar, which is shown in Figure 3. The parameter space below the solid line is excluded by the observations.

\hspace{-5mm}
\begin{figure}
\begin{center}
\includegraphics[scale=0.4]{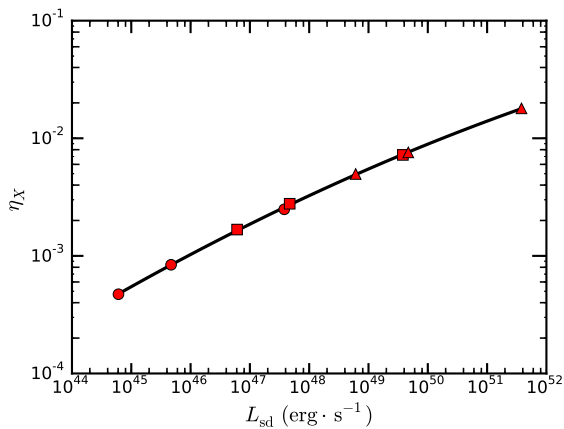}
\caption{Dependence of $\eta_X$ on the spin-down luminosity $L_{\rm sd}$. Different symbols are used to differentiate the surface magnetic field strength: circles, squares and triangles are for $B_s=10^{14},\,10^{15}$ and $10^{16}\,\rm G$ respectively.}
\end{center}
\end{figure}

\hspace{-5mm}
\begin{figure}
\begin{center}
\includegraphics[scale=0.4]{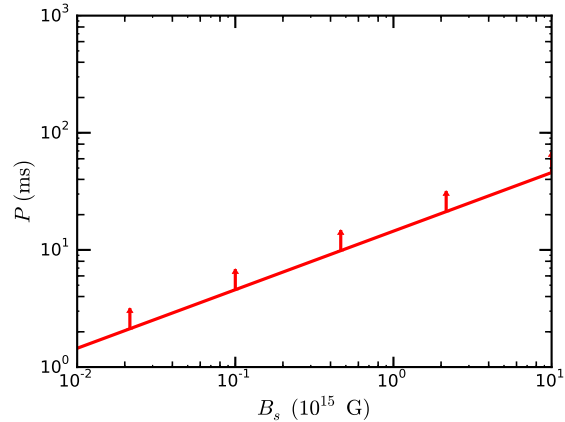}
\caption{Constraint on the spin period and surface magnetic field strength of a pulsar obtained from the requirement (\ref{eq:limit}). The reasonable parameter space lies above the solid line.}
\end{center}
\end{figure}

\hspace{-5mm}
\begin{figure}
\begin{center}
\includegraphics[scale=0.4]{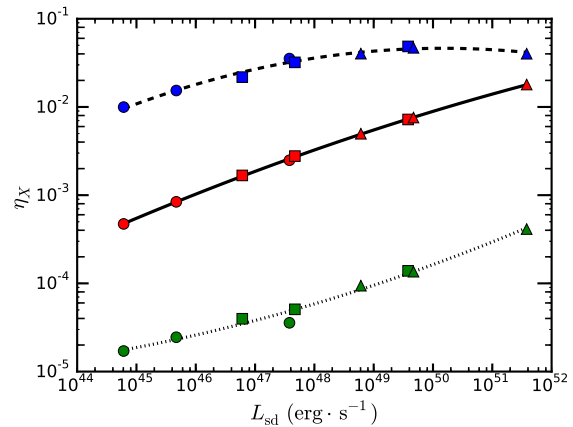}
\caption{Influence of different $\Gamma_{\rm sat}$ on the dependence of $\eta_X$ on $L_{\rm sd}$. The solid line is the same as in Figure 2, while the dashed line is for $\Gamma_{\rm sat}=10^3$ and the dotted line is for $\Gamma_{\rm sat}=10^5$.}
\end{center}
\end{figure}

\hspace{-5mm}
\begin{figure}
\begin{center}
\includegraphics[scale=0.4]{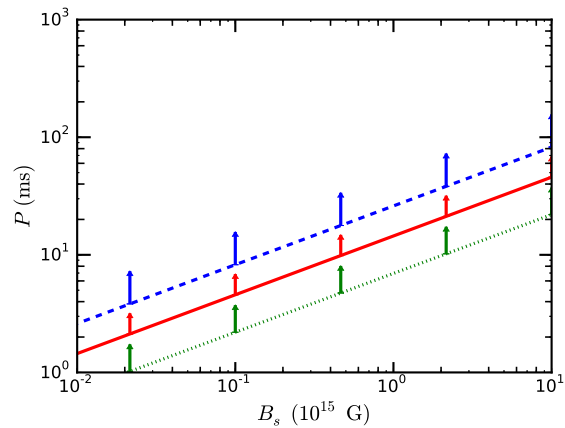}
\caption{Three constraints obtained for three values of $\Gamma_{\rm sat}$. The red solid line is the same as in Figure 3, while the blue dashed line is for $\Gamma_{\rm sat}=10^3$ and the green dotted line is for $\Gamma_{\rm sat}=10^5$.}
\end{center}
\end{figure}

We next consider the effect of the wind's saturation Lorentz factor ($\Gamma_{\rm sat}$) on the efficiency. To our knowledge, the saturation Lorentz factor $\Gamma_{\rm sat}$ depends on the initial magnetization parameter ($\sigma_0$) in the form of $\Gamma_{\rm sat}=\sigma_0^{3/2}$ \citep{dre02a, dre02b, gia05, ben17}. For a Poynting-flux-dominated outflow, the efficiency of converting magnetic energy to radiation is expected to rely on the magnetization parameter. This issue is worth investigating from a theoretical point of view, since it will help study the properties of a Poynting-flux-dominated outflow and then reveal the mystery of a central engine. In addition to the canonical $\Gamma_{\rm sat}=10^4$ assumed above, here we choose the other two values of $10^3$ and $10^5$ and thus new efficiencies are obtained. Strong dependence of $\eta_X$ on $\Gamma_{\rm sat}$ is shown in Figure 4, implying that lower magnetized outflows are more efficient at converting magnetic energy to radiation. The critical spin-down luminosities in the cases of $\Gamma_{\rm sat}=10^3$ and $10^5$ are $L_{\rm sd,cr}= 8.24\times10^{43}$ and $1.63\times10^{46}\,\rm erg\,s^{-1}$ respectively, and the corresponding constraints on $P$ and $B_s$ are shown in Figure 5. It is the $\Gamma_{\rm sat}=10^3$ case that places the most stringent constraint on the pulsar parameters.

\section{Conclusions and Discussion}
\label{sec:disc}
In this work we have assumed a rapidly rotating highly magnetized pulsar as the origin of FRB 121102 and constrained its spin period and magnetic field strength with upper limits given by current multi-wavelength observations. The magnetic energy dissipation in an isotropic pulsar wind should produce notable emission in the X-ray band. The non-detection by {\em XMM-Newton/Chandra} implies that the spin-down luminosity should be less than a critical value $L_{\rm sd,cr}$. We derived the efficiency of converting the spin-down luminosity to X-ray luminosity ($\eta_X$) and obtained its dependence on $L_{\rm sd}$. This efficiency depends strongly on the saturation Lorentz factor of the wind, or more intrinsically speaking, on the initial magnetization parameter of the wind. Outflows with a higher magnetization convert less energy to radiation. The reason for this is that the synchrotron emission turns from the fast-cooling regime to the slow-cooling regime as $\sigma_0$ increases, which is consistent with the conclusion by \citet{ben17}. Thus, for the three cases considered in this work, it is the $\Gamma_{\rm sat}=10^3$ case gives the most stringent constraint on the pulsar parameters. The method of using high-energy data to constrain some of the model parameters is relevant for newborn pulsars with ages younger than $T_{\rm sd}$, which is not the case for any known Galactic magnetar \citep{ten16}\footnote{Note that very bright, high-energy emission from internal magnetic dissipation in the wind cannot last for a very long time. In fact, its luminosity will decay significantly after the initial spin-down timescale of a newborn rapidly rotating highly magnetized pulsar. At later times, an additional dominant emission could arise from an interaction of the wind with its ambient medium or supernova ejecta \citep[for reviews see][]{gae06,sla17}.}. Also, the IGMD model is not easily tested with the current observations of pulsar wind nebulae (PWNe)\footnote{Actually, we have compared the IGMD model to observations of the present Crab nebula by taking the parameters of the Crab pulsar. With a distance of $2.2\,\rm kpc$, $B_s=4\times10^{12}\,\rm G$, and $P=33\,\rm ms$, the predicted flux ($\sim 10^{-11} \,\rm erg\,cm^{-2}\,s^{-1}$ at $5\,\rm keV$) of the high-energy emission from internal magnetic dissipation is far below the observed X-ray-to-gamma-ray flux of the Crab nebula \citep[$\sim 10^{-8} \,\rm erg\,cm^{-2}\,s^{-1}$ at $5\,\rm keV$; for a recent review see][]{buh14}. Therefore, the observed X-ray flux of the Crab nebula is dominated by the emission from shocks (in particular, a terminative reverse shock) produced by an interaction of the pulsar wind with its ambient gas.}. However, it will be testable with observations of GRBs or SLSNe, which are driven by newborn millisecond magnetars. In particular, this model will be possibly tested if the association of an FRB with a GRB or an SLSN is detected in the future.

We note that several works have placed limits on the pulsar scenario, but most of these limits were obtained from the observations of the persistent radio counterpart to FRB 121102. For instance, \citet{Kashiyama2017} studied the emission from a PWN in the framework of the ``burst-in-bubble'' model \citep{mur16}. With its application to the quasi-steady radio counterpart, they constrained the spin period and the magnetic field strength of the young pulsar by the minimum energy requirement for the PWN. The model in \citet{Dai2017} differs from \citet{Kashiyama2017} in a way that they considered a PWN without surrounding supernova ejecta and thus new constraints on the wind luminosity and the ambient medium density were obtained. Moreover, the age of the pulsar can be constrained by radio observations \citep{Bel2017, Metzger2017} and by other fair arguments like dispersion measures \citep{Cao2017, Kashiyama2017}. \citet{Lyutikov2017} argued that the energy source for FRB 121102 can also be constrained. In our paper, we focus on the X-ray-to-gamma-ray follow-up observations of FRB 121102 and our new constraints on $P$ and $B_s$ are generally consistent with previous works \citep{Cao2017, Dai2017, Kashiyama2017, Lyutikov2017}.

The difference between our work and \citet{zhabb17} lies not only in the selected upper limits, but also in the methods of calculation. Instead of simply assuming a constant radiation efficiency, we start from the realistic IGMD model of a pulsar wind and the efficiency is obtained in a more physical way. Determining the radiation efficiency has been a key issue for the Poynting-flux-dominated outflow in the previous studies, especially in the GRB field \citep[e.g.][]{ben15,ben16,ben17}. The method we developed in this work could be applied to various situations, and constraining the parameters of the pulsar origin of FRB 121102 is just one of them. More comprehensive work could be done with our method, such as applying it to short GRBs, magnetar giant flares, and even some black-hole accreting systems. These studies will appear elsewhere.

\acknowledgements
We thank Yun-Wei Yu, Bing Zhang, and an anonymous referee for helpful comments. This work was supported by the National Basic Research Program of China (973 Program grant 2014CB845800) and the National Natural Science Foundation of China grant 11573014.

\clearpage

\end{document}